# Multipole-phase beams: a new paradigm for structured waves

Gianluca Ruffato[1], Vincenzo Grillo[2], and Filippo Romanato[1,3]

[1]Department of Physics and Astronomy 'G. Galilei', University of Padova, via Marzolo 8, 35131 Padova, Italy

[2]CNR-Istituto Nanoscienze, Centro S3, Via G Campi 213/a, I-41125 Modena, Italy

[3]CNR-INFM TASC IOM National Laboratory, S.S. 14 Km 163.5, 34012 Basovizza, Trieste, Italy

Authors e-mails:

Gianluca Ruffato: gianluca.ruffato@unipd.it

Vincenzo Grillo: vincenzo.grillo@unimore.it

Filippo Romanato: filippo.romanato@unipd.it

Correspondence: Gianluca Ruffato, via Marzolo 8, Department of Physics and Astronomy 'G. Galilei', University of Padova, 35131 Padova (Italy), tel. +390498275933, fax. +390498277102

# ABSTRACT

The control of structured waves has recently opened innovative scenarios in the perspective of radiation propagation and light-matter interaction. In particular, the transmission of customized electromagnetic fields is investigated for telecommunications, with the aim of exploring new modulation formats besides the traditional, almost saturated, division multiplexing techniques. Beams carrying twisted wavefronts have long been recognized as the promising candidates, however their phase singularities and efficient multiplexing still raise open issues. In a more general insight into structured-phase beams, we introduce and develop here a new and unique paradigm based on the transmission of beams with harmonic phases having a multipole structure. The outlined framework encompasses multiplexing, transmission, and demultiplexing as a whole for the first time, describing wavefields evolution in terms of conformal mappings, and solving straightforwardly the critical issues of previous solutions. Because of its potentialities, versatility, and ease of implementation, we expect this completely new paradigm to find widespread applications for space division multiplexing especially in free space, from the optical to the microwave and radio regimes.

# INTRODUCTION

A deeper insight into the nature and properties of light has always inspired technological advances and disruptive applications. In fact, the exploitation of a particular property requires methods and components for the manipulation of that specific feature. With the flourishing of research and applications of structured light, the necessity to customize the intensity and phase distribution of a beam pursued the ability to process the wavefield analogously to spectral or polarisation decomposition. However, while the latter operations can be done straightforwardly, spatial decomposition is inherently more arduous to achieve and few practical examples exist. Some classes of beams have been identified in order to describe the proper method to perform spatial decomposition for specific transmission issues: Hermite- and Laguerre-Gaussian [1], Bessel [2], Elegant beams [3] are among the most famous. The exploration of these beam families and of their spatial degrees of freedom as control and manipulation parameters has paved the way to scientific milestones and innovative applications in a wide range of fields [4], in particular microscopy and imaging [5]-[7], micromanipulation [8][9], cryptography [10][11], and telecommunications [12][13].

In the telecom field, space division multiplexing (SDM) [14] relies on structuring the intensity or phase distribution of electromagnetic (EM) waves over a set of non-interfering spatial configurations to be used as distinct information channels at the same frequency in combination with standard modulation formats [15][16], offering a solution to the impelling problem of networks saturation [17]-[19], and a wider alphabet for quantum applications [20]. That requires the choice of a suitable family of orthogonal beams and the design of specific devices, i.e. the multiplexer and the demultiplexer, realizing their superposition and separation at the transmitter and at the receiver stage, respectively. The propagation in free space is mathematically expressed in terms of the Fourier transform [21], which allows to describe the evolution of a specific phase and intensity pattern from its generation up to detection. In the ray light picture, a point of the initial EM configuration is projected onto a point of the receiver area, realizing a sort of coordinate transformation of the transmitted field. It is somewhat surprising that, under reasonable assumptions of analyticity, the phase pattern Ω of the field must be harmonic, i.e. a solution of Laplace's equation in 2D [22].

During the last decade, most of the technological efforts of SDM have been focused on the exploitation of vortex beams [23] endowed with a phase term $\exp(i\ell\vartheta)$, being $\ell$ an integer unbounded value representing the amount of orbital angular momentum (OAM) [24][25] per photon in units of $\hbar$. The azimuthal phase gradient of OAM beams is indeed a particular solution of Laplace's equation, independent of the radial coordinate. While many significant milestones have been put in the generation and detection [26]-[37], manipulation [38]-[40], and propagation [41]-[43] of OAM beams, on the other hand their central phase singularity and divergence still represents a critical problem for long-distance transmission in free space [44]-[48], while the efficient generation and multiplexing with passive, versatile and all-optical few-elements method is arduous to achieve.

In a wider approach to the transmission of structured beams, we consider here the general solution of Laplace's equation for the phase, focusing on its properties in the view of space division multiplexing. A new set of beams is discovered, hereafter referred to as *multipole-phase beams*, devoid of central phase singularity and characterized by two continuous parameters, i.e. the phase strength and orientation. The same phase solutions are able to implement the mapping between multipole phases and linear momentum states, which can be easily processed with a Fourier lens, outlining an efficient method to achieve multiplexing and demultiplexing of those beams with only two optical elements.

The presented research paves the way to a novel and more general paradigm for enhanced-capacity transmission, hereafter called *multipole-phase division multiplexing* (Fig. 1). The underlying theory and experimental results lay the foundation, for the first time, of a complete and unique framework including as a whole the multiplexing, transmission and demultiplexing of spatial configurations of the electromagnetic fields, and it is expected to find promising applications especially for free-space transmission, from the optical up to the microwave and radio regimes.

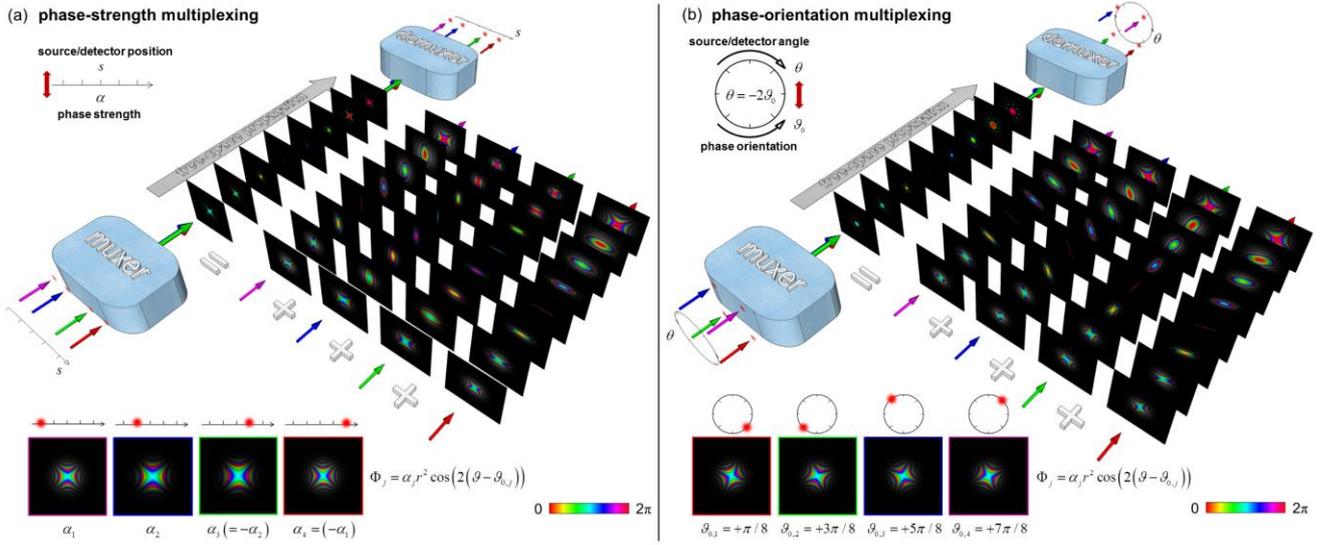

**Figure 1**. **Multipole-phase division multiplexing.** Scheme of multipole-phase division multiplexing working principle for the specific case of multipole order equal to +2. A linear (a) and a circular (b) distribution of independent Gaussian beams are first multiplexed into a set of superimposed beams with multipole phases, then demultiplexed after transmission. The two schemes can be used in combination and are here shown separately only for the sake of clarity. The multipole phase term of the *j*th beam is given by $\Phi_j=\alpha_j r^2\cos(2(\vartheta-\vartheta_{0,j}))$, defined by the phase strength $\alpha_j$ and orientation angle $\vartheta_{0,j}$. The modulation of these two degrees of freedom offers a 2D dense space $(\alpha_j, \vartheta_{0,j})$ over which the information channels can be multiplexed and demultiplexed at the same frequency. The (de)multiplexing devices allow to conjugate the two parameters to the axial displacement and orientation of the *j*th input/output beam, respectively. (a) Phase-strength multiplexing: beams carrying different phase strengths $\{\alpha_j\}$ are conjugated to distinct axial positions proportional to $\alpha_j$. (b) Phase-orientation multiplexing: beams with the same phase strength and different orientation angles $\{\vartheta_{0,j}\}$ are conjugated to distinct angular positions $\{-2\cdot\vartheta_{0,j}\}$ of the (de)multiplexed beams on the same circle. The intensity/phase patterns show the numerical propagation in free space over a length of 300 m at the wavelength of 632.8 nm. Brightness and colours refer to intensity and phase, respectively.

# RESULTS

**Theory and simulation**

As depicted in Fig. 1, the basic idea is to transform a linear or a circular distribution of (Gaussian) beams in a superposition of multipole-phase beams (multiplexing) and, after propagation, to perform the back transformation to the original distribution (demultiplexing). The phase of this new type of beams is described by two continuous parameters, defining the phase gradient and the angular orientation, which are put in one-to-one correspondence to the radial and angular positions of a 2D constellation of lasers and detectors, at the transmitter and the receiver sides, respectively.

In the paraxial regime, the propagation in free space of an initial field $U^{(i)} = u^{(i)}(r,\vartheta) e^{i\Omega(r,\vartheta)}$ is described by the Fresnel diffraction integral [49]. The stationary phase approximation provides a useful tool to solve an integral of that type, and it is equivalent, from a physical point of view, to a ray-light approach and eventually to a coordinate transformation $\boldsymbol{\rho} \equiv \boldsymbol{\rho}(r,\vartheta)$, induced by the phase term $\Omega$ [50][51]. In the quest of a phase structure to be exploited for space division multiplexing, we consider a phase pattern imparting a transformation which is conformal. Under this constrain, it is straightforward to show (see Supplementary Information S1) that the phase $\Omega$ must be harmonic, i.e. that it satisfies Laplace's equation in 2D, whose general solution under variable separation is given by:

$$\Omega(r,\vartheta) = \alpha r^m \cos\left(m(\vartheta - \vartheta_0)\right) \tag{1}$$

Due to the analogy with the integrated fields of multipoles in electrostatics [52], we will refer to a phase pattern as that in Eq. (1) as a multipole phase, being $\alpha$ and $\vartheta_0$ continuous parameters describing the phase strength and orientation, respectively. The beams families are identified by the parameter $m$ that measures the multiplicity order of the phase. For $m > 0$, the phase in Eq. (1) is defined over the whole plane $(r,\vartheta)$, therefore it does not induce points of null intensity associated to phase singularities or discontinuities. In general, under free-space propagation, a field with multipole phase of order $m$ is Fourier transformed into a multipole phase of order $m/(m-1)$, with phase strength and orientation depending on the input values (see Supplementary Information).

With the aim to exploit those beams for structured light applications, it becomes essential to find out, for each order $m$, an effective method to generate and sort the multipole phases on the basis of their defining parameters $\alpha$ and $\vartheta_0$. The idea is to perform a mapping between multipole phases and linear phase gradients, which can be easily multiplexed or separated by means of a Fourier lens. This is achieved by implementing an $n$-fold circular-sector transformation [38] with factor $n = -1/m$, which maps conformally a point $(r, \vartheta)$ to the new polar coordinates $(\rho, \varphi) \equiv \left(a(r/b)^{-1/n}, \vartheta/n\right) = \left(a(r/b)^m, -m\vartheta\right)$, being $a$ and $b$ arbitrary scaling parameters. The transformation performs substantially a scaling on the azimuthal coordinate, while the power scaling on the radial coordinate is dictated by the Cauchy-Riemann conditions that a conformal mapping must satisfy. Under this transformation, the phase in Eq. (1) is transformed into a linear phase gradient:

$$\Phi^{out}(\rho, \varphi) = \beta \rho \cos(\varphi - \varphi_0) \tag{2}$$

with the definitions $\beta = \alpha b^m / a$ and $\varphi_0 = -m\vartheta_0$. As shown in [38][39] (see Supplementary File S2 for detailed calculations), a circular-sector transformation with factor $n = -1/m$ can be imparted by a phase-only element with transmission function equal to $\exp(i\Omega^D_{m,1})$, where

$$\Omega^D_{m,1}(r, \vartheta) = K \cdot r^{1+m} \cos((1+m)\vartheta) \tag{3}$$

where $K = kab^{-m}/(f(1+m))$. Actually, to complete the phase transformation a further phase plate is required in $z = f$, accounting for the phase distortions due to the propagation. This second element exhibits a phase $\Omega^D_{m,2}$ with the same functional dependence as that in Eq. (3), under the substitutions

$m \to 1/m$ and $a \leftrightarrow b$ (plus a quadratic focusing to introduce a Fresnel correction). As a matter of fact, this element performs the inverse transformation when the optical system is illuminated in reverse, i.e. a circular-sector transformation by a scaling factor $1/n$. Comparing Eq. (3) and Eq. (1), it is worth noting that the two elements required to process the multipole order $m$, are endowed with multipole phases as well, of orders $1+m$ and $1+1/m$, respectively.

After placing a lens in cascade to the second element in *f-f* configuration, a bright spot is expected on its back focal plane at the polar coordinates:

$$(R,\theta) = \left(\frac{f_F}{k}\beta, \varphi_0\right) = \left(\frac{f_F}{k}\frac{\alpha b^m}{a}, -m\vartheta_0\right) \quad (4)$$

being $f_F$ the focal length of the Fourier lens. Therefore, beams endowed with multipole phases of the same order $m$, but with different strength and orientation values, can be mapped onto distinct points in far field by using the sequence of an $n$-fold circular-sector transformation with $n=-1/m$ and a Fourier lens.

Conversely, the same configuration can be used to multiplex several input beams into a collimated bunch of multipole phases with the same order $m$ but different phase strengths and orientations. Actually, the first element is not exactly as the second one in the demultiplexer. As a matter of fact, since the inverse transformation performs the mapping of the whole pattern onto a circular sector with amplitude $2\pi/|m|$, an $|m|$-fold multiplication is needed in order to obtain a beam defined over the whole $2\pi$ range. Therefore, the required phase turns out to be the combination of $|m|$ phase patterns performing $m$-fold circular-sector transformations, rotated of $2\pi/|m|$ with respect to each other (Supplementary Information S3).

In order to prove the working principle of the beams transformation, we will focus our analysis on the specific case $m=+2$. The importance of this order is dictated by its invariance under Fourier transform, i.e. free-space propagation, since it satisfies condition $m = m/(m-1)$. However, this choice allows to describe the characteristic properties of multipole-phase beams without any loss of generality. For the sake of clarity, we will focus on the demultiplexing stage, due to the symmetry of the inverse multiplexing process under time reversal.

The demultiplexing scheme is depicted in Fig. 2, where the input field is mapped into a linear phase gradient by means of a circular-sector transformation with $n=-1/2$ and sorted by a Fourier lens in cascade. The phase patterns of the required optical elements are obtained by substituting $m=+2$ and $m=+1/2$ in Eq. (3), plus quadratic terms for focusing and Fresnel correction. Their phase patterns are shown in Figs. 2(b) and 2(c) (without the quadratic terms). A numerical simulation of the propagation from the first phase plate up to the second one is reported (Fig. 2(d.1-8)). Input beams with different phase strengths and orientation angles are expected to generate linear phase gradients with different components of the wavevector on the plane, being focused at distinct positions on the back-focal plane of a lens. As shown in Fig. 3 for different phase strengths and orientations in input, the far-field spots are clearly distinguishable, allowing to detect the input beams and measure their carried energy. In Supplementary Information S3 a design and simulation of the multiplexing device is reported.

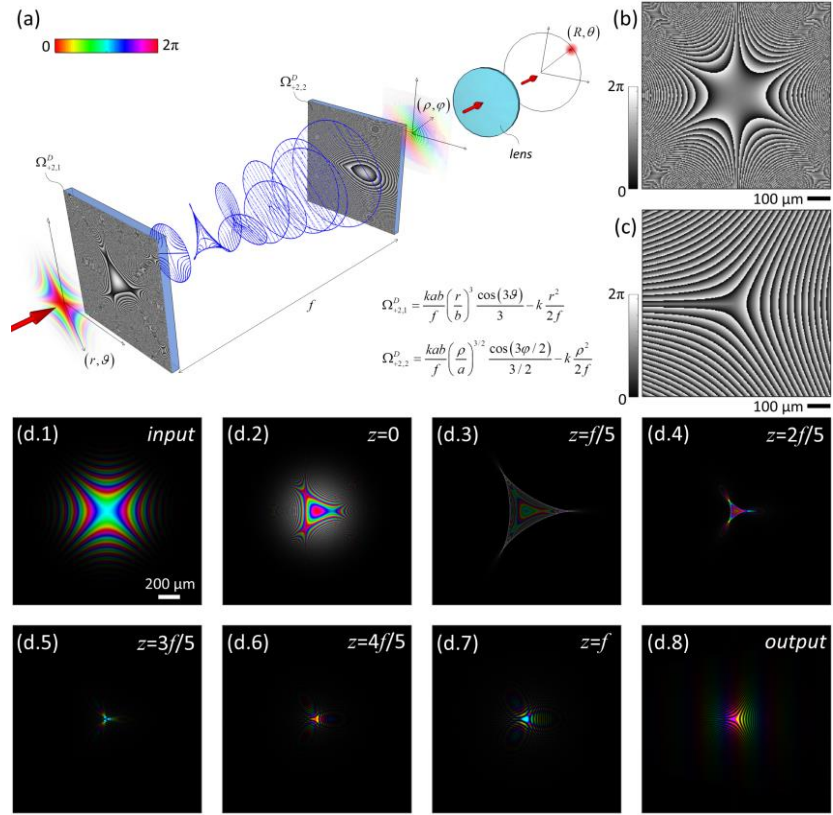

**Figure 2**. **Transformation of a multipole-phase beam of order $m=+2$.** (a) Scheme: the input field is transformed by the first phase element imparting a circular-sector transformation with $n=-1/2$ and a final linear phase gradient is retained after the second phase element (performing the inverse circular sector transformation, i.e. $n=-2$). A lens is used for Fourier transform. Phase patterns of the first (b) and second (c) phase element, without the quadratic focusing (only the circular-sector transformation phase contribution is shown). Design parameters: $a=500$ μm, $b=300$ μm, $f=10$ mm, $\lambda=632.8$ nm, input beam waist $w_0=312.5$ μm. (d) Numerical simulation of the propagation of an input Gaussian beam carrying a multipole phase with $m=+2$, $\alpha=1.0\cdot10^{-4}$ μm$^{-2}$ (d.1) at different position on the $z$-axis, after illuminating the first element (d.2), up to the second optical element (d.3-d.7, $f/5$ step), and output phase-corrected beam (d.8). Colours and brightness refer to phase and intensity, respectively.

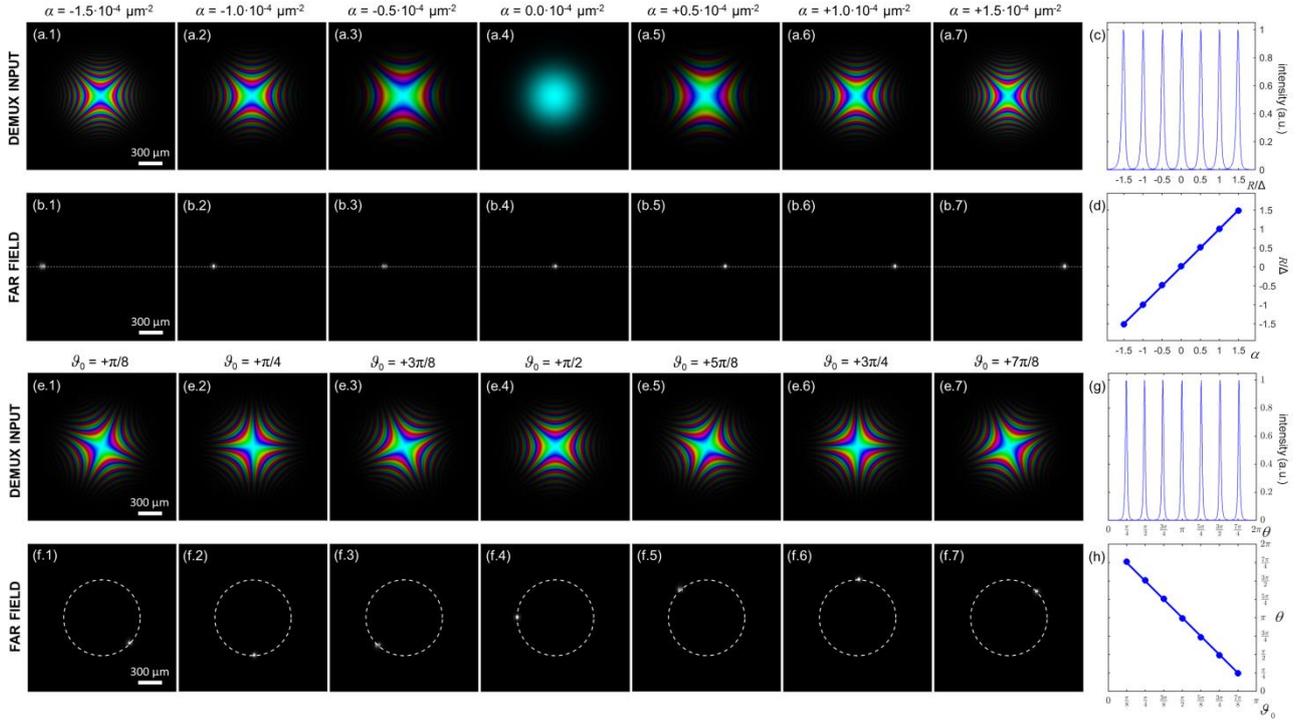

**Figure 3. Sorting of multipole-phase beams of order $m=+2$.** Numerical simulations of different input beams with a multipole phase of order $m=+2$ entering the optical scheme depicted in Fig. 2. After applying the circular-sector transformation with $n=-1/2$, the multipole phase $\Phi_{2,j}^{in}(r,\vartheta) = \alpha_j r^2 \cos(2(\vartheta - \vartheta_{0,j}))$ of the $j$th beam is transformed into a linear phase $\Phi_{2,j}^{out}(\rho,\varphi) = \beta_j \rho \cos(\varphi - \varphi_{0,j})$, where $\beta_j = \alpha_j b^2/a$ and $\varphi_{0,j} = -2\vartheta_{0,j}$, and mapped to a point at the position $(R_j, \theta_j) = (f_F/k \cdot \beta_j, -2\vartheta_{0,j})$ at the back focal plane of a lens with focal length $f_F$. Design parameters of the transformation: $a$=500 μm, $b$=300 μm, $f$=10 mm, $\lambda$=632.8 nm, input waist $w_0$=312.5 μm. Focal length for far-field analysis: $f_F$=20 cm. (a.1-7) The beams are endowed with different strength parameter $\alpha$ in the set $\{0, \pm 0.5, \pm 1.0, \pm 1.5\} \cdot 10^{-4}$ μm$^{-2}$. As expected, a bright spot appears in the far field (b.1-7), at a position proportional to the input phase strength $\alpha$. (c) Far-field cross-section on the white dashed line in (b). (d) Normalized spots position $R/\Delta$ ($\Delta = f_F/k \cdot b^2/a$) and theoretical trend (blue solid line) according to Eq. (4), as a function of the phase strength $\alpha$ [$10^{-4}$ μm$^{-2}$]. (e.1-7) The beams are endowed with the same strength parameter $\alpha$=+1.0·10$^{-4}$ μm$^{-2}$ and different rotation angles $\vartheta_0$ from +π/8 to +7π/8, step +π/8. As expected, the bright spots appear over the same circle in far field at different angular positions equal to -2$\vartheta_0$ (f.1-7). (g) Far-field cross-section on the white dashed circle in (f). (h) Spots angular position $\theta$ and theoretical trend (blue solid line), as a function of the rotation angle $\vartheta_0$.

**Experimental tests**

The possibility to measure optically the parameters of a multipole phase has been verified experimentally for illumination under beams of order $m=+2$ at the wavelength of 632.8 nm, generated by a reflective Liquid-Crystal-on-Silicon (LCoS) spatial light modulator (SLM) using a phase/amplitude modulation technique [53] (Fig. 4(a)). The transformation optics imparting a circular-sector transformation with $n=-1/2$ was implemented by means of a second LCoS SLM, using the two halves of the display to upload the two distinct phase patterns, i.e. transformer and phase-corrector, and deploying a mirror for back-reflection as shown in Fig. 4(b, c). This configuration reduces the degrees of freedom and simplifies the alignment procedure, as already shown in experimental implementations with diffractive optics [34][54]. This is achieved by adding a phase term $\gamma_1 \cdot x$ to the first pattern, and centring the second element, i.e. the phase-corrector, at a distance $\gamma_1 f/k$ from the centre of the first. Then, an additional spatial frequency carrier $\gamma_2=1.5 \cdot \gamma_1$ was added to the phase-corrector, in order to tilt the output beam out of its optical axis, as shown in the scheme in Fig. 4(a). In the specific, the experimental parameters chosen for the circular-sector transformation were $a=1000$ μm, $b=1500$ μm, and $f=30$ cm. A lens term with focal length $f_F=40$ cm was added to the phase-corrector pattern.

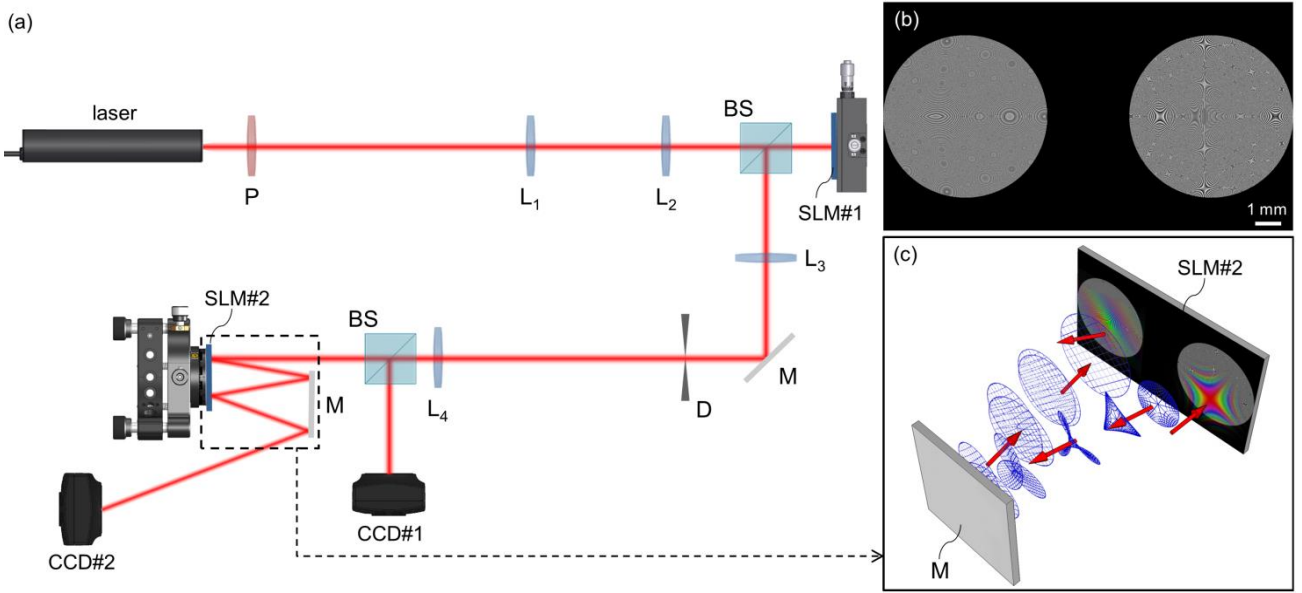

**Figure 4**. **Scheme of the experimental setup for optical tests.** (a) The laser beam ($\lambda$=632.8 nm, waist size 240 μm) is linearly polarized (P) and expanded ($f_1$=2.54 cm, $f_2$=15.0 cm). A beam splitter (BS) is used to extract the back-reflected beam after phase/intensity reshaping with a first SLM (SLM#1) for multipole-phase generation. The beam is filtered (D) and resized ($f_3$=20.0 cm, $f_4$=25.0 cm) before illuminating a second SLM (SLM#2) on its first half implementing the circular-sector transformer. A mirror (M) is used for back-reflection on the second half of the display for phase correction (scheme in (c)). (b) Phase pattern uploaded on SLM#2: 1920 x 1080 pixels, pixel size 8 μm x 8 μm, 256 grey levels. A second beam splitter is used to check the beam intensity distribution on a first camera CCD#1. A second camera CCD#2 is placed on the focal plane of the phase-correcting pattern ($f_F$=40.0 cm).

As discussed in ref. [32] for the *log-pol* case, the phase plates are calculated in the stationary phase approximation for planar wavefronts, then they are expected to work perfectly only for collimated beams in input. Irrespective of their divergence, beams carrying multipole phases are not plane waves, and the local deviation due to the structured phase front is equal to $|\nabla\Phi^{(in)}|/k$. However, the optical transformation still works as expected, provided that the angular deviation introduced by the transformer dominates over any input deviation from normal incidence. For a given multipole order $m$, it is required that $|\nabla\Phi^{(in)}| \ll |\nabla\Omega_{1,m}^D|$, providing the condition $\alpha \ll kaw_0/(fmb^m)$, where $w_0$ is the input waist (Supplementary Material S4). For $m=+2$, we find $\alpha \ll kaw_0/(2fb^2)$, defining an upper threshold to the input strength that the system can measure without dramatic deviations.

The capability of the optical configuration to measure strength and orientation angle of multipole phases of order $m=+2$ was experimentally tested for input beams with a Gaussian intensity profile and a beam waist $w_0=2.5$ mm. With the given parameters of the optical transformation and of the input beam, the upper threshold for the measurable phase strength resulted around $8 \cdot 10^{-5}$ μm$^{-2}$, however significant deviations started to appear for values close to $2 \cdot 10^{-5}$ μm$^{-2}$. Therefore, we limited the analysis up to a maximum value of $1.5 \cdot 10^{-5}$ μm$^{-2}$. In Fig. 5, the output is reported for different values of the phase strength $\alpha$ at fixed orientation $\vartheta_0=0$. As expected, the far-field spots are arranged over the same line and they shift proportionally to $\alpha$. Their positions exhibit a linear trend as a function of the input strength, and the value of the experimental slope $\Delta_{exp}=+(20.20\pm0.08)\cdot 10^{+6}$ μm$^3$ is in good accordance with the expected theoretical value $\Delta_{th}=f_L b^2/(ka)=+20.14\cdot 10^{+6}$ μm$^3$. Then, we considered input beams with fixed multipole-phase strength $\alpha=+10\cdot 10^{-6}$ μm$^{-2}$ and varying rotation angle $\vartheta_0$, increasing from $\vartheta_0=0$ at steps of $\Delta\vartheta_0=\pi/8$. As shown in Figs. 6(a), the angular position $\theta$ of the output spots varies according to $-2\vartheta_0$, as expected.

These results suggest the possibility to perform spatial-division multiplexing over a selected range of values of phase strengths and orientation angles. The values should be chosen properly in order to minimize the superposition of the corresponding spots in the far field, i.e. to reduce to cross-talk between the channels. In Fig. 6, experimental data are shown regarding the demultiplexing of several superpositions of multipole-phase beams endowed with the same phase strength $\alpha=+10\cdot 10^{-6}$ μm$^{-2}$ but different orientation angles in the set $\{\vartheta_{0,j}=(j-1)\pi/8\}$, $j=1,\ldots,8$. As expected, the optical system is able to sort the different contributions according to the parameters of the carried multipole phases.

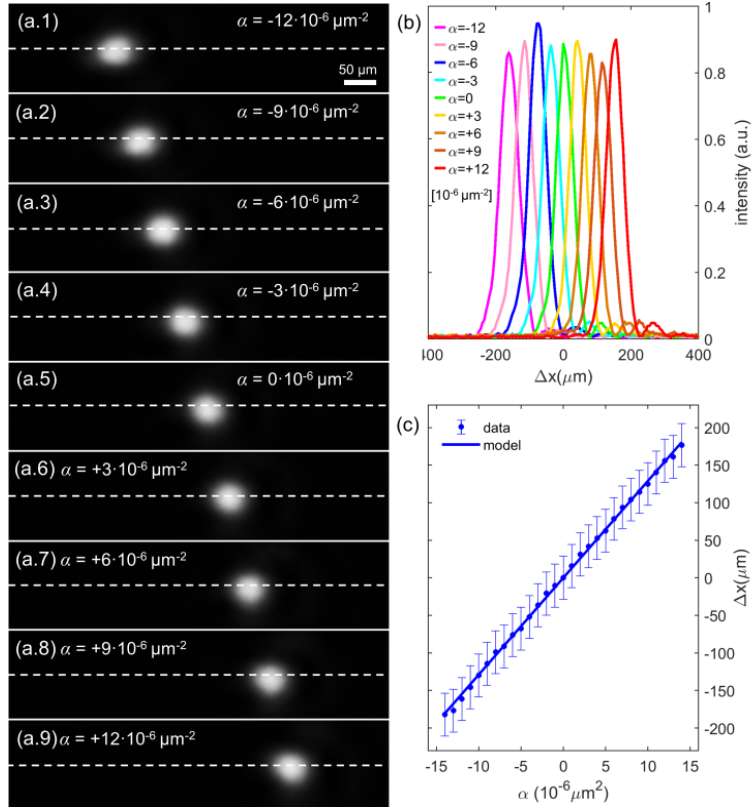

**Figure 5**. **Measurement of multipole-phase strength, $m=+2$.** (a.1-9) Experimental output for varying multipole-phase strength $\alpha$ in the range from $-12\cdot 10^{-6}$ μm$^{-2}$ (a.1) to $+12\cdot 10^{-6}$ μm$^{-2}$ (a.9), step $3\cdot 10^{-6}$ μm$^{-2}$, fixed orientation $\vartheta_0=0$, and cross-sections (b) on the white dashed lines. (c) Far-field position as a function of the input strength $\alpha$ in the range from $-15\cdot 10^{-6}$ μm$^{-2}$ to $+15\cdot 10^{-6}$ μm$^{-2}$, step $1\cdot 10^{-6}$ μm$^{-2}$: experimental data (blue dots) and theoretical model (blue line). Error bar: half-width half-maximum. As expected, the shift of the far-field spot is proportional to the input strength $\alpha$. Optical processing with $n=-1/2$ circular-sector transformation, design parameters: $a=2000$ μm, $b=1000$ μm, $f=30$ cm, $f_F=40$ cm, $\lambda=632.8$ nm. Theoretical slope: $\Delta_{th}=f_L b^2/(ka)=+20.14\cdot 10^{+6}$ μm$^3$. Experimental slope: $\Delta_{exp}=+(20.20\pm 0.08)\cdot 10^{+6}$ μm$^3$.

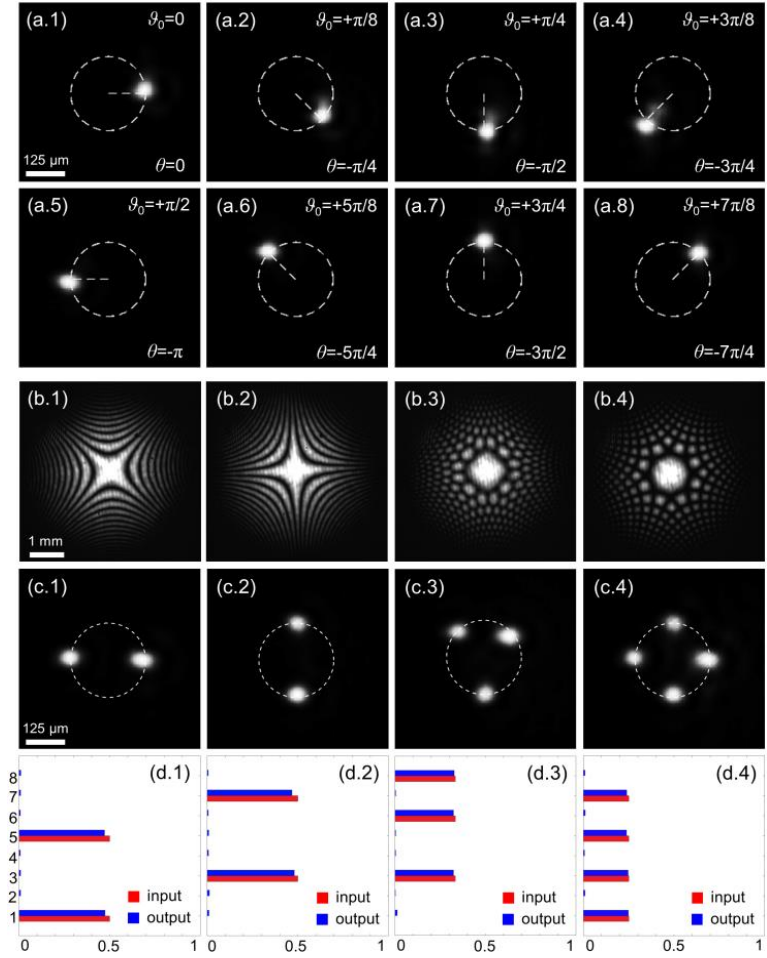

**Figure 6**. **Measurement and demultiplexing of multipole-phase orientation angle, $m=+2$.** (a.1-8) Experimental output for varying multipole-phase orientation $\vartheta_0$ in the range of 8 values from $\vartheta_0=0$ to $\vartheta_0=7\pi/8$, step $\pi/8$. Fixed phase strength $\alpha=+10\cdot10^{-6}$ µm$^{-2}$. Optical processing with $n=-1/2$ circular-sector transformation, design parameters: $a=2000$ µm, $b=1000$ µm, $f=30$ cm, $f_F=40$ cm, $\lambda=632.8$ nm. As expected, the output spots appear at the angular position $\theta=-2\vartheta_0$. The same set of beams has been used to test the demultiplexing capability of the system. (b.1-4) Intensity of the generated beams superposition, with orientation angles $\vartheta_0=0$ & $\vartheta_0=\pi/2$ (b.1), $\vartheta_0=\pi/4$ & $\vartheta_0=3\pi/4$ (b.2), $\vartheta_0=\pi/4$ & $\vartheta_0=5\pi/8$ & $\vartheta_0=7\pi/8$ (b.3), $\vartheta_0=0$ & $\vartheta_0=\pi/4$ & $\vartheta_0=\pi/2$ & $\vartheta_0=3\pi/4$ (b.4). (c.1-4) Experimental output after the demultiplexer. (d.1-4) Comparison between input (in red) and output (in blue) energy distributions over the selected range of channels. The $j$th channel corresponds to the input rotation angle $\vartheta_{0,j}=(j-1)\pi/8$. The total power has been normalized to unity.

## DISCUSSION AND CONCLUSIONS

In a more general approach to the propagation of structured-phase beams, we developed here an innovative framework based on the transmission of completely new beams carrying multipole harmonic phases. In the landscape of structured light, those beams present unique possibilities in terms of compact all-optical multiplexing and sorting, solving straightforwardly the open issues of previous techniques. As a matter of fact, the steps of generation, propagation, and detection are here considered as a whole for the first time in terms of conformal transformations of the wavefields, and the common ground is represented by the condition of harmonicity that the phases must satisfy. Although they are not modes in the strict sense, the existence of quasi-orthogonal multipole-phase patterns which are invariant under Fourier transform suggests the possibility to exploit those beams as distinct information carriers for space division multiplexing.

For a given multipole order $m$, the two continuous parameters of phase strength and orientation angle define a 2D dense state space over which a set of quasi-orthogonal multipole-phase beams can be defined to carry distinct information channels over the same frequency. Although in principle any value of $m$ can be chosen, the case $m=+2$ represents the ideal case for transmission due to the invariance of the multipole order under Fourier transform, and the consequent symmetry between the multiplexing and demultiplexing devices. The core of those devices is represented by a conformal mapping between multipole phases and linear phase gradients, implemented by means of two confocal phase plates realizing a circular-sector transformation of factor $n=-1/m$. In the framework of optical vortices, those transformations were introduced to perform the multiplication and division of optical OAM in an efficient and compact manner [39]. In the present context, the same mapping has been applied for the measurement of structured beams endowed with multipole phases, representing a special and peculiar class of phase patterns with respect to the abovementioned transformation. While the sorting has been demonstrated using computer-controlled SLMs, the phase elements can be fabricated as diffractive optics [54] or in a metasurface form [22][55] and integrated into compact architectures, scaling straightforwardly to the more compact sizes considered in the simulation section.

Conformal transformations are an effective tool for wavefront manipulation and require only two confocal phase plates to perform a unitary transformation of the intensity and phase distributions of the input field. Thus, this method provides an all-optical solution for beam shaping which is more compact, feasible and efficient than other techniques based for instance on holograms [26], on the cascade of many optical operations [27]-[29] or on integrated plasmonic or photonic platforms [30]. The key-idea is to map conformally the structured wavefields into linear momentum states which can be generated or separated straightforwardly by using a Fourier lens. In the context of OAM beams, this has been performed by unwrapping the azimuthal phase gradient into a linear one using the *log-pol* mapping, which since its first implementation with SLM in 2010 [31], has known increasing improvements in miniaturization [36], resolution [34] and integration [54][55]. However, the axial-symmetry breaking and diffraction limitations still prevent a compact and efficient multiplexing of OAM beams, and additional optical elements are required for beam reshaping before and after the multiplexing and demultiplexing stages [36][37]. Moreover, the (de)multiplexing can be easily implemented only on one integer parameter, i.e. the azimuthal index $\ell$, then the input/output channels are arranged along a fixed line.

With respect to the *log-pol* sorter for OAM beams, the technique here presented outlines a more versatile implementation of the (de)multiplexing devices and opens to a wider range of degrees of freedom. In fact, the parameters used to define a multipole phase, i.e. its strength $\alpha$ and rotation angle $\vartheta_0$ on the plane, are continuous variables, and the only constraint on their choice is to avoid the overlapping of the corresponding bright spots at the detector stage, i.e. to guarantee a maximum threshold of cross-talk among them. In addition, the angular degree of freedom allows to cover the whole plane and to distribute the channels over a two-dimensional space. Furthermore, the output spots present a Gaussian-like intensity profile, avoiding the need for cumbersome beam reshaping.

This is even more advantageous in the multiplexer design since the output beam of a standard source can illuminate the optics directly (Supplementary Information S3).

For positive multipole orders, the phase term is devoid of phase singularity, then it does not induce points of null intensity on the carrying beams, differently from the azimuthal phase of optical vortices. In those beams in fact, the out-of-axis component of the Poynting vector leads to a divergence of the vortex and to a consequent expansion of its dark centre, which increases with the OAM value, affecting the efficiency of the receiver and limiting the application to short distances [43]-[46]. In the Supplementary Information, the simulated propagation of a multipole-phase beam with Gaussian intensity distribution is reported.

As far as guided propagation is concerned, beams with multipole phases are not exact solutions of the wave equation in the medium, then they should be properly described in terms of superposition of the fibre modes, with a consequent distortion of the input phase and intensity distributions due to the modal dispersion. On the other hand, due to the low modal dispersion and the self-imaging properties, graded-index (GRIN) fibres are expected to represent the best candidate for multipole-phase beams transmission [56]. For this reason, an immediate application of this spatial division multiplexing technique could be found in the field of data centres, where GRIN fibers are already deployed and the maximum distance length is around a few kilometres.

It is worth noting that the particular case of multipole phases of order $m=+2$ resembles the pattern of the well-known 2-fold astigmatisms. Then, the present method offers a solution to measure the amount and orientation of astigmatisms in a fast and direct way, and analogous schemes can be designed for higher order aberrations described by multipole phases with positive integer $m$ ($m$-fold astigmatism)[57].

We envisage that the present work paves the way to a completely new SDM paradigm: the multipole-phase division multiplexing (MPDM). The optical schemes presented in this study outline the key configurations for the design of the multiplexing and demultiplexing devices required to implement a communication link, while the information carriers are provided by a defined set of multipole phases of the same order but different strength and orientation values. The axial centred intensity makes this multiplexing configuration particularly efficient for free-space point-to-point communication extending its application beyond the optical wavelengths. The exploitation of SDM has been acquiring increasing attention in the microwave range for next-generation free-space networks [58], where the realization of diffractive optics [59] and metasurfaces [60] for the control of the phase structure has recently been shown. Therefore, we envisage the implementation of communication links in free-space based on a new generation of microwave diffractive antennas for the transmission and detection of multipole-phase beams in the microwave regime.

# METHODS

## Numerical simulations

Custom codes were developed in MatLab environment, based on the convolution algorithm applied to the Fresnel diffraction integral [61], in order to simulate the multiplexing and demultiplexing of beams endowed with multipole phases. The phase patterns were defined over a square mesh of 5633 x 5633 pixels with a spatial resolution of 312.5 nm. Simulations have been performed for several phase strengths and orientation angles of multipole-phase beams of order $m=+2$, demonstrating the possibility to generate and sort this kind of beams using a circular-sector transformation with $n=-1/2$.

## Optical characterization

Multipole-phase beams were generated using a first LCoS spatial light modulator (SLM) (PLUTO-NIR-010-A, Holoeye) by applying a phase and amplitude modulation technique [52]. The SLM display was illuminated by a Gaussian beam ($\lambda= 632.8$ nm, beam waist $w_0=240$ μm, power 0.8 mW) emitted by a HeNe laser (HNR008R, Thorlabs), linearly polarized and expanded using a first telescope ($f_1=2.54$ cm, $f_2=15.0$ cm). A second telescope ($f_3=20.0$ cm, $f_4=25.0$ cm) was placed after the first SLM in order to isolate and image the first-order encoded mode onto a second LCoS SLM display (PLUTO-NIR-010-A, Holoeye) mounted on a 6-axis kinematic mount (K6XS, Thorlabs). A 50:50 beam-splitter was used to split the beam and check the input beam profile with a first CMOS camera (DCC1545M, Thorlabs). The multipole-phase beam illuminated the first zone of the SLM display, performing the $n$-fold circular-sector transformation, then it was back-reflected by a mirror onto the second half implementing phase-correction. The design parameters chosen for the circular-sector transformation were: $a=1000$ μm, $b=1500$ μm, $f=30$ cm. The mirror was placed on a kinematic mount (KM100, Thorlabs) and its distance from the SLM display, equal to half the focal length of the first phase pattern, i.e. 15 cm, could be finely controlled with a micrometric translator (TADC-651,

Optosigma). The SLM display was formed by 1920 x 1080 pixels of size 8 μm x 8 μm. Each phase pattern was designed inside a circle of radius $L/5=3.072$ mm, and the distance between the two centres was set to $3L/5=9.216$ mm, being $L$ the width of the SLM display. Finally, the transformed beam was collected by a second CMOS camera (DCC1545M, Thorlabs), placed on the focus of the phase-corrector pattern ($f_F=40$ cm).

# ACKNOWNLEDGEMENTS

This work was supported by project VORTEX 3 from CEPOLISPE. V.G. acknowledges Q-SORT, a project funded by the European Union's Horizon 2020 Research and Innovation Program under grant agreement No. 766970.

# CONFLICTS OF INTEREST

The authors declare no competing financial interests.

# AUTHOR CONTRIBUTIONS

G.R. developed the theoretical background, performed the design and numerical simulations, and conducted the experimental tests. V.G. contributed to the theory and design. F.R. addressed the aims of the project and managed the laboratory. All authors discussed and contributed to the writing of the manuscript.

# SUPPLEMENTARY MATERIAL

## Multipole-phase beams: a new paradigm for structured waves

Gianluca Ruffato[1], Vincenzo Grillo[2], and Filippo Romanato[1,3]

[1]Department of Physics and Astronomy 'G. Galilei', University of Padova, via Marzolo 8, 35131 Padova, Italy

[2]CNR-Istituto Nanoscienze, Centro S3, Via G Campi 213/a, I-41125 Modena, Italy

[3]CNR-INFM TASC IOM National Laboratory, S.S. 14 Km 163.5, 34012 Basovizza, Trieste, Italy

Authors e-mails:

Gianluca Ruffato: gianluca.ruffato@unipd.it

Vincenzo Grillo: vincenzo.grillo@unimore.it

Filippo Romanato: filippo.romanato@unipd.it

Correspondence: Gianluca Ruffato, via Marzolo 8, Department of Physics and Astronomy 'G. Galilei', University of Padova, 35131 Padova (Italy), tel. +390498275933, fax. +390498277102

## S1. LAPLACE'S EQUATION IN STRUCTURED PHASE TRANSMISSION

In the paraxial approximation, the propagation of a structured field $U^{(i)}(r,\vartheta) = u^{(i)}(r,\vartheta)e^{i\Omega(r,\vartheta)}$ is described by the diffraction integral:

$$U(\rho,\varphi,z) = \frac{1}{i\lambda z}\iint u^{(i)}(r,\vartheta)e^{i\Omega(r,\vartheta)}e^{-ik\frac{r\rho}{z}\cos(\vartheta-\varphi)}rdrd\vartheta \tag{1}$$

According to the stationary phase approximation [1,2] a two-dimensional integral with the form in Eq. (1) can be approximated with its contributions around the saddle points of the total phase function $\Phi$ of the argument, that is:

$$\Phi(r,\vartheta) = \Omega(r,\vartheta) - k\frac{r\rho}{z}\cos(\vartheta-\varphi) \tag{2}$$

and Eq. (1) can be rewritten in the form:

$$U(\rho,\varphi,z) \cong \frac{2\pi\sigma}{\lambda z}\frac{u^{(i)}(r^*,\vartheta^*)}{\sqrt{H}}e^{i\Phi(r^*,\vartheta^*)} \tag{3}$$

where $H$ is the Hessian determinant of $\Phi$, $\sigma = \text{sgn}(\partial^2\Phi/\partial x^2)$ when $H>0$, $\sigma = -i$ otherwise, $(r^*,\vartheta^*)$ are the saddle points of $\Phi$, that is the solutions of the differential equation $\nabla\Phi = 0$, where the differential operator should be intended in 2D. Then, the field in $z$ is endowed with a phase function $\Omega^{(z)}$ given by:

$$\Omega^{(z)}(\rho,\varphi) = \Phi(r^*,\vartheta^*) \tag{4}$$

As a consequence of the relation in Eq. (2), the condition $\nabla\Phi = 0$ leads to the identity:

$$\nabla\Omega = \frac{k}{z}\boldsymbol{\rho} \tag{5}$$

where $\boldsymbol{\rho} = \rho(\cos\varphi,\sin\varphi) \equiv (u,v)$ in Cartesian coordinates. Under the existence of partial derivatives for $\boldsymbol{\rho}$, Eq. (5) implies the following relation:

$$\frac{\partial u}{\partial y} = \frac{\partial v}{\partial x} \tag{6}$$

Eq. (5) establishes a connection between a point $\boldsymbol{r} = (r,\vartheta)$ of the input phase and a point $\boldsymbol{\rho} = (\rho,\varphi)$ on the plane in $z$, describing a sort of optical transformation, imparted by the phase $\Omega$, of the input intensity pattern $u^{(i)}(r,\vartheta)$. With the aim of exploiting the phase pattern to transfer information, an assumption on this transformation is that it locally conserves the angles, i.e. it is conformal. Using the complex formalism, a conformal mapping of the point $\zeta = x+iy$ satisfying the condition in Eq. (6) is expressed by an anti-holomorphic function $g(\overline{\zeta}) = u(x,y)+iv(x,y)$. In

addition to Eq. (6), the anti-holomorphic mapping satisfies the following Cauchy-Riemann condition:

$$\frac{\partial u}{\partial x} = -\frac{\partial v}{\partial y} \tag{7}$$

Taking the divergence in 2D of Eq. (5) and using Eq. (7) we obtain the following condition on the phase:

$$\nabla^2 \Omega = 0 \tag{8}$$

It should be noted that the previous condition is still valid in the Fresnel regime, by including a focusing term $\exp(-ikr^2/2f)$ and considering the propagated field at $z=f$.

The explicit form of Eq. (8) in polar coordinates is given by:

$$\left( \frac{1}{r}\frac{\partial}{\partial r} r \frac{\partial}{\partial r} + \frac{1}{r^2}\frac{\partial^2}{\partial \vartheta^2} \right) \Omega = 0 \tag{9}$$

Solving Eq.(9) under variable separation gives the following general solution:

$$\Omega(r,\vartheta) = \alpha r^m \cos(m(\vartheta - \vartheta_0)) \tag{10}$$

where *m* is assumed to be integer, after imposing periodic boundary conditions in $\vartheta = 0$. The solution with $\sin(\cdot)$ can be chosen without loss of generality. Trivial solutions of Eq. (8) are given by $\Omega(r) = \alpha \log(r/b)$, which is axially symmetric, and $\Omega(\vartheta) = m\vartheta$, independent on the radial coordinate. The latter recalls the azimuthal phase term which is peculiar of orbital angular momentum (OAM) beams. Therefore, in this general approach to structured-phase division multiplexing in the paraxial regime, OAM beams are rediscovered as particular solutions of the phase patterns which can be transferred between two parties.

## S2. CIRCULAR-SECTOR TRANSFORMATION IN THE STATIONARY PHASE APPROXIMATION

For the benefit of the reader, we provide here a detailed calculation of the phase elements required to optically perform an *n*-fold circular-sector transformation. In the paraxial approximation, the propagation of an input beam $U^{(i)}$ at a distance *z*, after illuminating a phase plate with transmission function $exp(i\Omega)$ located at $z=0$, is described by the Fresnel diffraction integral:

$$U(\rho,\varphi,z) = \frac{e^{ik\frac{\rho^2}{2z}}}{i\lambda z} \iint U^{(i)}(r,\vartheta) e^{i\Omega(r,\vartheta)} e^{ik\frac{r^2}{2z}} e^{-ik\frac{r\rho}{z}\cos(\vartheta-\varphi)} r dr d\vartheta \tag{11}$$

According to the stationary phase approximation [1,2] the integral can be approximated with its contributions around the saddle points of the phase function. If we consider the phase function Ω in Eq. (11) as the sum of two contributions, i.e. the phase term $\Omega_n$ imparting the desired *n*-fold circular-sector transformation and a quadratic focusing term:

$$\Omega(r,\vartheta)=\Omega_n(r,\vartheta)-k\frac{r^2}{2f} \qquad (12)$$

then the nullification of the gradient of the total phase inside the integral in Eq.(11) leads to the following condition:

$$\nabla\Omega_n=\frac{k}{z}\boldsymbol{\rho}-\frac{k}{z}\left(1-\frac{z}{f}\right)\mathbf{r} \qquad (13)$$

We now consider the coordinate change:

$$(\rho,\varphi)=\left(a\left(\frac{r}{b}\right)^{-\frac{1}{n}},\frac{\vartheta}{n}\right) \qquad (14)$$

describing a mapping between the reference frame $(r,\vartheta)$ on the input plane $(z=0)$ and the new reference frame $(\rho,\varphi)$ on the destination plane at $z=f$. This mapping performs a rescaling of the azimuthal angle by a factor $n$, while the power scaling on the radial coordinate is dictated by the Cauchy-Riemann conditions [3]. After substituting Eqs. (14) in Eq. (13) and solving for $z=f$, we obtain:

$$\nabla\Omega_n=\frac{k}{f}a\left(\frac{r}{b}\right)^{-\frac{1}{n}}\left(\cos\left(\frac{\vartheta}{n}\right),\sin\left(\frac{\vartheta}{n}\right)\right) \qquad (15)$$

The integration can be done easily, remembering the definition:

$$\nabla\Omega_n\cdot\vec{r}=\frac{\partial\Omega_n}{\partial r} \qquad (16)$$

being $\vec{r}=(\cos\vartheta,\sin\vartheta)$. We obtain:

$$\frac{\partial\Omega_n}{\partial r}=k\frac{a}{f}\left(\frac{r}{b}\right)^{-\frac{1}{n}}\cos\left(\frac{\vartheta}{n}\right)\cos(\vartheta)+k\frac{a}{f}\left(\frac{r}{b}\right)^{-\frac{1}{n}}\sin\left(\frac{\vartheta}{n}\right)\sin(\vartheta)=k\frac{a}{f}\left(\frac{r}{b}\right)^{-\frac{1}{n}}\cos\left(\vartheta-\frac{\vartheta}{n}\right) \qquad (17)$$

After a straightforward integration we get the final result:

$$\Omega_n(r,\vartheta)=k\frac{ab}{f}\left(\frac{r}{b}\right)^{1-\frac{1}{n}}\frac{\cos\left(\left(1-\frac{1}{n}\right)\vartheta\right)}{1-\frac{1}{n}} \qquad (18)$$

Actually, a second phase plate is required in $z=f$ in order to complete the phase transformation and account for the phase distortions due to propagation. In the reverse configuration, this element works as a circular-sector transformation with a factor $1/n$, performing the inverse optical transformation:

$$(r,\vartheta) = \left(b\left(\frac{\rho}{a}\right)^{-n}, n\varphi\right) \tag{19}$$

After performing analogous calculations as above, we find out the following phase pattern for the phase-corrector element:

$$\Omega_{n.PC}(\rho,\varphi) = k\frac{ab}{f}\left(\frac{\rho}{a}\right)^{1-n}\frac{\cos((1-n)\varphi)}{1-n} \tag{20}$$

which shows basically the same expression as in Eq. (18), under the substitutions $b \leftrightarrow a$, $n \rightarrow 1/n$ and $(r,\vartheta) \rightarrow (\rho,\varphi)$.

**S3. MULTIPLEXER DESIGN**

Under the substitution $n=-1/m$, Eq. (18) and Eq. (20) provide the phase patterns of the sequence of optical elements, i.e. transformer and phase-corrector, required to demultiplex a superposition of multipole-phase beams of order $m$:

$$\Omega^{D}_{m,1}(r,\vartheta) = \frac{2\pi ab}{\lambda f}\left(\frac{r}{b}\right)^{1+m}\frac{\cos((1+m)\vartheta)}{1+m} \tag{21}$$

$$\Omega^{D}_{m,2}(\rho,\varphi) = \frac{2\pi ab}{\lambda f}\left(\frac{\rho}{a}\right)^{1+\frac{1}{m}}\frac{\cos\left(\left(1+\frac{1}{m}\right)\varphi\right)}{1+\frac{1}{m}} \tag{22}$$

When illuminated in reverse, the same configuration can be exploited to multiplex several input beams into a collimated bunch of multipole-phase beams with the same order but differing in phase strength and rotation angle on a plane perpendicular to the propagation direction. However, the first element is not exactly as the second one in the demultiplexer (Eq. (22)). Since the inverse transformation performs the mapping of the whole pattern onto a circular sector with amplitude $2\pi/|m|$, then an |m|-fold multiplication of the input beam is needed in order to obtain a beam defined over the whole 2π range. For this reason, the required phase turns out to be the combination of $|m|$ phase patterns performing $n$-fold circular-sector transformations with $n=-m$ and rotated by $2\pi/|m|$ with respect to each other. The phase pattern of the first element, i.e. the multiplier, is described by the combination:

$$\Omega^{M}_{m,1}(r,\vartheta) = \arg\left\{\sum_{p=1}^{|m|}e^{i\Omega^{D,(p)}_{m,2}}\right\} \tag{23}$$

where:

$$\Omega_{m,2}^{D,(p)}(r,\vartheta) = \frac{2\pi ab}{\lambda f}\left(\frac{r}{a}\right)^{1+\frac{1}{m}} \frac{\cos\left[\vartheta\left(1+\frac{1}{m}\right)+(p-1)\frac{2\pi}{|m|}\right]}{1+\frac{1}{m}} \quad (24)$$

while the second element of the multiplexer is given again by Eq. (21), that is $\Omega_{m,2}^{M} = \Omega_{m,1}^{D}$.

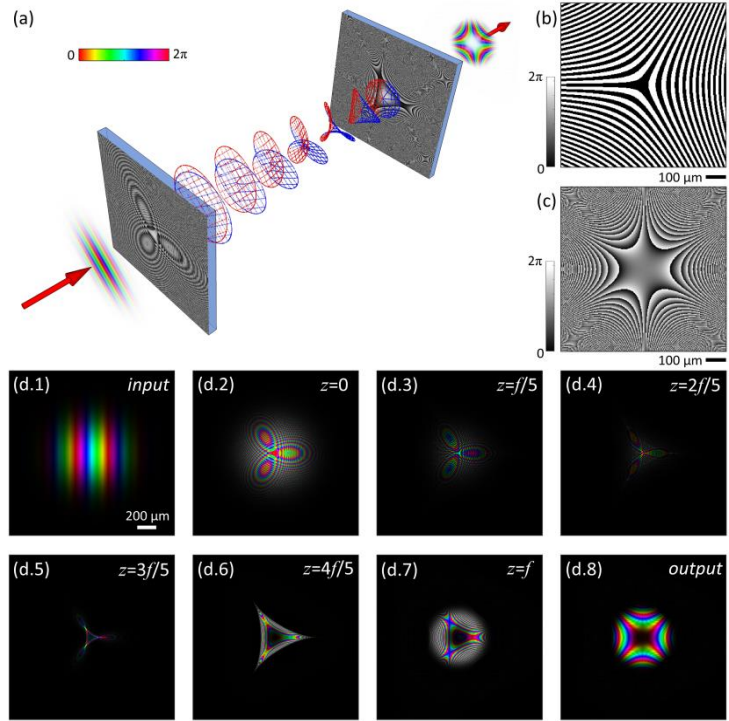

**Figure S1. Generation of multipole-phase beams of order *m*=+2.** (a) Scheme: the input linear phase is transformed by the first phase element imparting a multiplication by a factor *n*=-2, then a final multipole phase of order *m*=+2 is retained after the second phase element (performing a circular sector transformation with *n*=-1/2). Phase patterns of the first (b) and second (c) phase elements, without the focusing term (only the transformation phase pattern is shown). Design parameters: *a*=500 μm, *b*=300 μm, *f*=10 mm, *λ*=632.8 nm. (d) Numerical simulations of the propagation of an input tilted Gaussian beam (d.1) at different position on the *z*-axis, after illuminating the first element (d.2), up to the second optical element (d.3-d.7, *f*/5 step), and output phase-corrected beam (d.8). Colours and brightness refer to phase and intensity, respectively.

In Fig. S1 a scheme of the optical configuration for the generation of a beam carrying a multipole-phase term with *m*=+2 is depicted. An input linear phase gradient is transformed into the desired phase by applying a 2-fold multiplication plus a phase inversion (*n*=-2). While the second element (Fig. S1(c)) is equal to the first element of the demultiplexer, in this case the first element (Fig. S1(b)) is given by the combination of two phase elements rotated by π, as a result of Eqs. (23) and (24). The strength and orientation of the output phase can be controlled by acting properly on the input linear phase gradient. As shown in Fig. S2 and Fig. S4, supposing it is generated by illuminating an *f-f* optical system with a Gaussian beam, it is possible to tune the output phase strength *α* and its rotation angle $\vartheta_0$ by changing the axial displacement of the input beam according to:

$$\begin{cases} \vartheta_0 = -\dfrac{\theta}{2} \\ \alpha = \dfrac{k}{f_F}\dfrac{aR}{b^2} \end{cases} \quad (25)$$

$(\theta, R)$ being the polar coordinates of the input beam position on the first focal plane of the lens and $f_F$ its focal length.

The overlap between two output fields $\psi_{\alpha_i,\vartheta_{0,i}}$ and $\psi_{\alpha_j,\vartheta_{0,j}}$ with different phase parameters can be calculated with the integral:

$$w_{ij} = \frac{\left|\left\langle \psi_{\alpha_i,\vartheta_{0,i}} \middle| \psi_{\alpha_j,\vartheta_{0,j}} \right\rangle\right|^2}{\left\langle \psi_{\alpha_i,\vartheta_{0,i}} \middle| \psi_{\alpha_i,\vartheta_{0,i}} \right\rangle \left\langle \psi_{\alpha_j,\vartheta_{0,j}} \middle| \psi_{\alpha_j,\vartheta_{0,j}} \right\rangle} \quad (26)$$

And the cross-talk on the $i$th channel can be evaluated by:

$$XT_i = -10\log\left(\frac{\sum_{j\neq i} w_{ji}}{\sum_{j} w_{ji}}\right) \quad (27)$$

As shown in Fig. S3 and S5 for the two sets of beams generated in Fig. S2 and S4, respectively, the output beams can be assumed to be quasi-orthogonal, due to the negligible cross-talk among them. As a matter of fact, since there is no spatial superposition between the input spots illuminating the *f-f* system, that null overlap is expected to be maintained also during beam multiplexing and propagation, due to the unitary nature of the conformal mapping and free-space propagation.

As shown in Fig. S2 and S4, the intensity distribution exiting the multiplexer exhibits a decrease in intensity close the centre. if the multipole phase is uploaded onto a Gaussian intensity distribution, this choice increases the cross-talk, as shown in Fig. S5, however its values, below -40 dB, can be still acceptable for telecom applications.

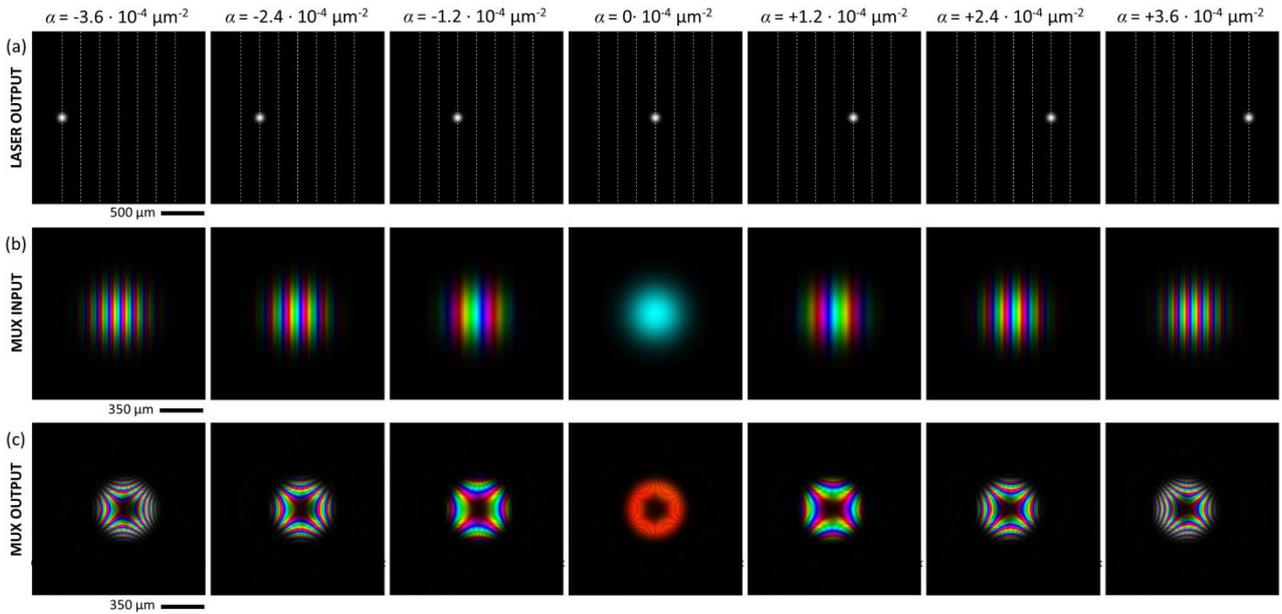

**Figure S2. Generation of multipole-phase beams of order *m*=+2 with different strength *α*.** Numerical simulation of the generation of different beams carrying multipole phases of order *m*=+2 using the optical scheme depicted in Fig. S1. A Gaussian spot (a) is Fourier transformed at the back-focal plane of lens ($f_F$=10 cm) into a Gaussian beam with tilt angle (b) depending on the axial displacement of the input beam. The scheme in Fig. S1 maps the input linear phase into an output multipole phase of order *m*=+2 (c). Design parameters: *a*=500 µm, *b*=300 µm, *f*=10 mm, $\lambda$=632.8 nm. By controlling the axial displacement of the input beam (a) it is possible to tune the strength *α* of the output multipole phase (c), according to Eqs. (25).

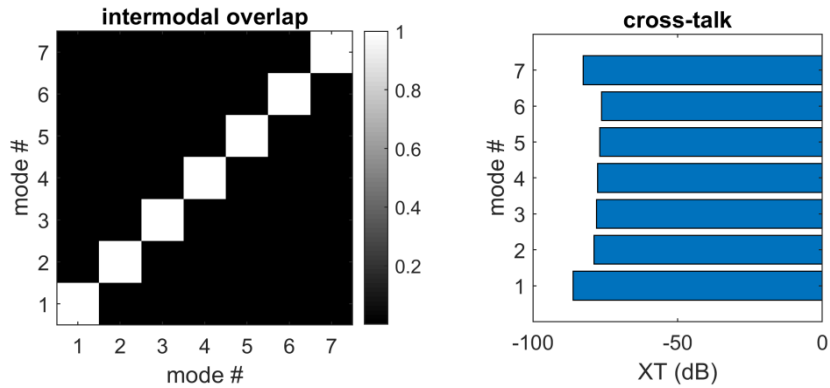

**Figure S3. Intermodal overlap and cross-talk of the beams generated in Fig. S2.**

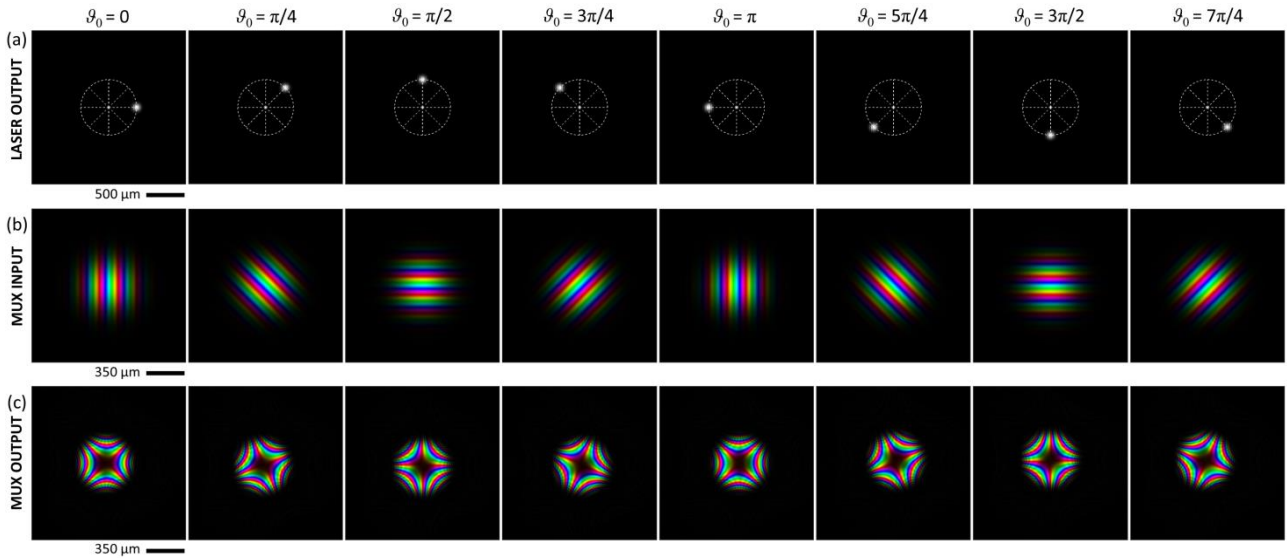

**Figure S4. Generation of multipole-phase beams of order *m*=+2 with different orientation angle.** Numerical simulation of the generation of different beams carrying multipole phases of order *m*=+2 using the optical scheme depicted in Fig. S1. A Gaussian spot (a) is Fourier transformed at the back-focal plane of lens ($f_F$=10 cm) into a Gaussian beam with azimuthal orientation (b) depending on the angular position of the input beam. Fixed phase strength $\alpha=2\cdot10^{-4}$ µm$^{-2}$. Design parameters as in Fig. S2.

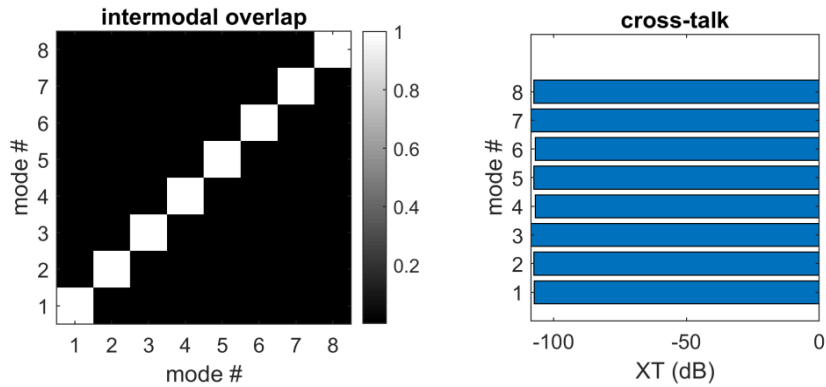

**Figure S5. Intermodal overlap and cross-talk of the beams generated in Fig. S4.**

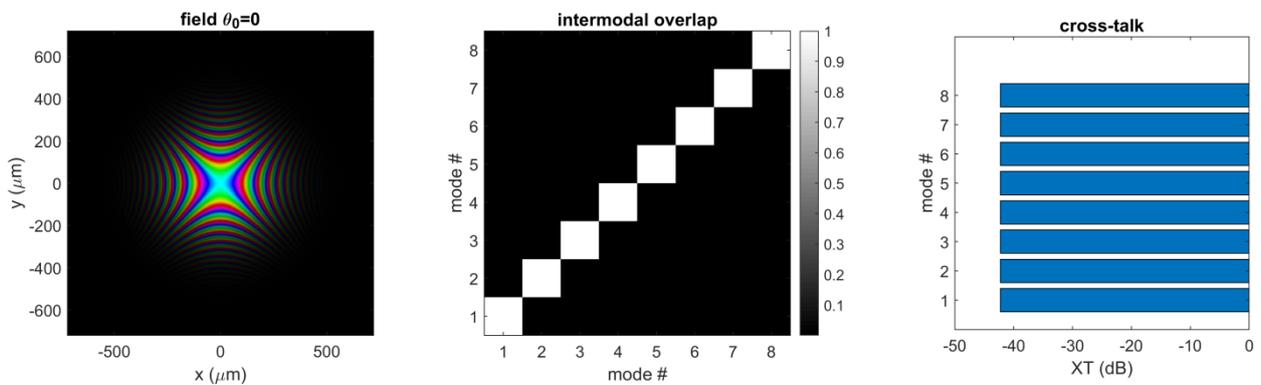

**Figure S6. Intermodal overlap and cross-talk of multipole-phase beams with a Gaussian intensity profile and varying orientation angle as in Fig. S4.**

# S4. DISTORTION OF CIRCULAR-SECTOR TRANSFORMATION FOR MULTIPOLE PHASES IN INPUT

When the stationary phase approximation is applied, it is usually assumed that the input field impinges on the first phase element with a planar wavefront. On the other hand, a multipole-phase beam illuminates the optics with a phase contribution $\Phi^{(in)}(r,\vartheta) = \alpha r^m \cos(m(\vartheta - \vartheta_0))$. The effect of a non-planar input wavefronts has been for instance discussed in the case of *log-pol* optical transformation [4], when the twisted wavefront introduces a distortion in the output field distribution that is negligible as far as the input OAM is far below a threshold value. The same analysis has been performed for the circular-sector transformation applied to OAM multiplication and division [3], and an analogous limit for the input OAM has been calculated, depending on the transformation parameters and on the size of the input beam. In the following, we apply the same analysis to the case of a multipole phase in input. In the paraxial ray approximation, a ray passing through a phase mask $\Omega$ at the position $(x_0, y_0)$, placed in the plane $z=0$, is deflected at an angle $1/k \cdot \nabla \Omega|_{x_0,y_0}$. Then, the intersection with a plane placed in $z$ has the following coordinates [5]:

$$r(z) = r_0 + \frac{z}{k}\nabla\Omega\Big|_{r_0} \tag{28}$$

If a multipole-phase beam illuminates the optical element with a phase function $\Omega_n$ given by Eq. (18), we have to include in Eq. (28) the contribution of the initial phase gradient, that is $\Omega = \Omega_n + \Phi^{(in)}$. Therefore, on a plane perpendicular to the propagation direction at the position $z$, the different rays emerging from a radial position $r$ of the phase mask in Eq. (12), intersect at the following points, describing the parametric curve:

$$\begin{cases} x(z) = r\cos\vartheta\left(1 - \frac{z}{f}\right) + \frac{az}{f}\left(\frac{r}{b}\right)^{-\frac{1}{n}}\cos\left(\frac{\vartheta}{n}\right) + \frac{z}{k}\alpha m r^{m-1}\cos((m-1)\vartheta - m\vartheta_0) \\ y(z) = r\sin\vartheta\left(1 - \frac{z}{f}\right) + \frac{az}{f}\left(\frac{r}{b}\right)^{-\frac{1}{n}}\sin\left(\frac{\vartheta}{n}\right) - \frac{z}{k}\alpha m r^{m-1}\sin((m-1)\vartheta - m\vartheta_0) \end{cases} \tag{29}$$

where we used the definition $(x_0, y_0) = (r\cos\vartheta, r\sin\vartheta)$. At the focal plane $z = f$, we obtain:

$$\begin{cases} x(f) = a\left(\frac{r}{b}\right)^{-\frac{1}{n}}\cos\left(\frac{\vartheta}{n}\right) + \frac{f}{k}\alpha m r^{m-1}\cos((m-1)\vartheta - m\vartheta_0) \\ y(f) = a\left(\frac{r}{b}\right)^{-\frac{1}{n}}\sin\left(\frac{\vartheta}{n}\right) - \frac{f}{k}\alpha m r^{m-1}\sin((m-1)\vartheta - m\vartheta_0) \end{cases} \tag{30}$$

Considering the radial distance $R = \sqrt{x^2 + y^2}$ as a function of the angle $\varphi$ on the focal plane, and recalling the relation $\varphi = \vartheta/n$, we obtain:

$$R(r,\varphi) = \sqrt{a^2\left(\frac{r}{b}\right)^{-\frac{2}{n}} + \left(\frac{f}{k}\alpha m r^{m-1}\right)^2 + 2\frac{af\alpha m}{kb^{-1/n}}r^{m-\frac{1}{n}-1}\cos\left[n(m-1)\varphi - m\vartheta_0 + \varphi\right]} \tag{31}$$

If the circular-sector transformer has been chosen in order to perform the sorting of the input multipole phase, then the relation $n=-1/m$ holds, and the previous equation becomes:

$$R(r,\varphi) = \sqrt{a^2\left(\frac{r}{b}\right)^{2m} + \left(\frac{f}{k}\alpha m r^{m-1}\right)^2 + 2\frac{af\alpha m}{kb^m}r^{2m-1}\cos\left[\frac{\varphi}{m} - m\vartheta_0\right]} \tag{32}$$

Therefore, a periodic oscillation is introduced, with a period equal to $2\pi|m|$. However, the distortion introduced by the input wavefront is negligible, provided the following condition is satisfied:

$$\frac{f}{k}\alpha m r^{m-1} \ll a\left(\frac{r}{b}\right)^m \tag{33}$$

which suggests a condition on the input phase strength $\alpha$:

$$\alpha \ll \frac{kaw_0}{fmb^m} \tag{34}$$

where we considered the beam waist $w_0$ as estimation for the average radius of the input beam.

## S5. FREE-SPACE PROPAGATION

In this section we analyse the propagation in free-space of a beam endowed with a multipole-phase term of order $m$:

$$U^{(i)}(r,\vartheta) = u^{(i)}(r)e^{i\alpha_0 r^m \cos(m(\vartheta-\vartheta_0))} \tag{35}$$

where $\alpha_0$ is the initial multipole-phase strength, $u^{(i)}(r)$ is an axially symmetric field distribution. With the aim of transmitting up to a distance $f$, we impart a focusing term $\exp(-ikr^2/2f)$ to the beam. The field at a distance $z$ is described by the diffraction integral:

$$U(\rho,\varphi,z) = \frac{e^{ik\frac{\rho^2}{2z}}}{i\lambda z}\iint u^{(i)}(r)e^{i\alpha_0 r^m \cos(m(\vartheta-\vartheta_0))}e^{-ik\frac{r^2}{2f}}e^{ik\frac{r^2}{2z}}e^{-ik\frac{r\rho}{z}\cos(\vartheta-\varphi)}rdrd\vartheta \tag{36}$$

In particular, at $z=f$ we have:

$$U(\rho,\varphi,f) = \frac{e^{ik\frac{\rho^2}{2f}}}{i\lambda f}\iint u^{(i)}(r)e^{i\alpha_0 r^m \cos(m(\vartheta-\vartheta_0))}e^{-ik\frac{r\rho}{f}\cos(\vartheta-\varphi)}rdrd\vartheta \tag{37}$$

By applying the stationary phase approximation, the following relation holds (see Eq. (3)):

$$U(\rho,\varphi,f) \propto \frac{e^{ik\frac{\rho^2}{2f}}}{i\lambda f}\frac{u^{(i)}(r^*)}{\sqrt{H}}e^{i\Phi(r^*,\vartheta^*)} \tag{38}$$

where $(r^*, \vartheta^*)$ are the stationary points and $H$ is the Hessian determinant of the phase term $\Phi$ of the argument, that is:

$$\Phi(r,\vartheta) = \alpha_0 r^m \cos(m(\vartheta - \vartheta_0)) - k\frac{\rho r}{f}\cos(\vartheta - \varphi) \tag{39}$$

The condition $\nabla \Phi = 0$ is equivalent to the following equations in polar coordinates:

$$\begin{aligned} \frac{\partial \Phi}{\partial r} &= m\alpha_0 r^{m-1} \cos(m(\vartheta - \vartheta_0)) - k\frac{\rho}{f}\cos(\vartheta - \varphi) = 0 \\ \frac{\partial \Phi}{\partial \vartheta} &= -m\alpha_0 r^m \sin(m(\vartheta - \vartheta_0)) + k\frac{\rho r}{f}\sin(\vartheta - \varphi) = 0 \end{aligned} \tag{40}$$

The solution of the previous system is straightforward and leads to:

$$\begin{cases} \rho = \dfrac{mf\alpha_0}{k} r^{m-1} \\ \varphi = (1-m)\vartheta + m\vartheta_0 \end{cases} \tag{41}$$

After inverting Eqs. (41) and substituting in (38) we obtain the phase term of the beam in Eq. (37) after propagation:

$$\begin{aligned} \Omega^{(f)}(\rho,\varphi) = \Phi(r^*,\vartheta^*) &= \alpha_0 \left(\frac{k\rho}{mf\alpha_0}\right)^{\frac{m}{m-1}} \cos\left(m\left(\frac{\varphi}{1-m} - \frac{m}{1-m}\vartheta_0 - \vartheta_0\right)\right) - \\ &\quad - \frac{k\rho}{f}\left(\frac{k\rho}{mf\alpha_0}\right)^{\frac{1}{m-1}} \cos\left(\frac{\varphi}{1-m} - \frac{m}{1-m}\vartheta_0 - \varphi\right) = \\ &= \alpha_0(1-m)\left(\frac{k\rho}{mf\alpha_0}\right)^{\frac{m}{m-1}} \cos\left(\frac{m}{m-1}(\varphi - \vartheta_0)\right) \end{aligned} \tag{42}$$

As a consequence, the transmitted beam is still endowed with a multipole phase term, defined by a new order equal to $m' = m/(m-1)$. The condition $m' = m$ is satisfied for the trivial case $m = 0$, and for $m = 2$.

This result makes definitely clear the peculiar role played by multipole-phase beams of order $m = 2$. As a matter of fact, their phase structure is preserved under Fourier transform, therefore they represent the best candidate to transmit information in free-space in the paraxial regime. The same consideration is expected to be valid for propagation in graded-index fibers, due to their low modal dispersion and the self-imaging property.

As suggested by Eq. (38), a quadratic term $\exp(-ik\rho^2/2f)$ should be included also at the receiver in order to compensate for the Fresnel term. Then, for $m = 2$, the transmitted phase $\Omega^{(i)}(r,\vartheta) = \alpha_0 r^2 \cos(2(\vartheta - \vartheta_0))$ is transformed into a multipole-phase of the same order $\Omega^{(f)}(\rho,\varphi) = \alpha_1 \rho^2 \cos(2(\varphi - \vartheta_1))$, where:

$$\alpha_1 = \frac{k^2}{4f^2} \frac{1}{\alpha_0} \tag{43}$$

and $\vartheta_1 = \vartheta_0 + \pi/2$ (where the $\pi/2$ term arises from the factor $1-m=-1$ in Eq. (42)). If a $2f$ system is put in cascade, by applying Eq. (37) again we finally have a phase strength $\alpha_2$ which is proportional to the input initial one and given by:

$$\alpha_2 = \frac{\alpha_0}{M^2} \tag{44}$$

where $M = f_2/f$ is the magnification factor of the whole optical system.

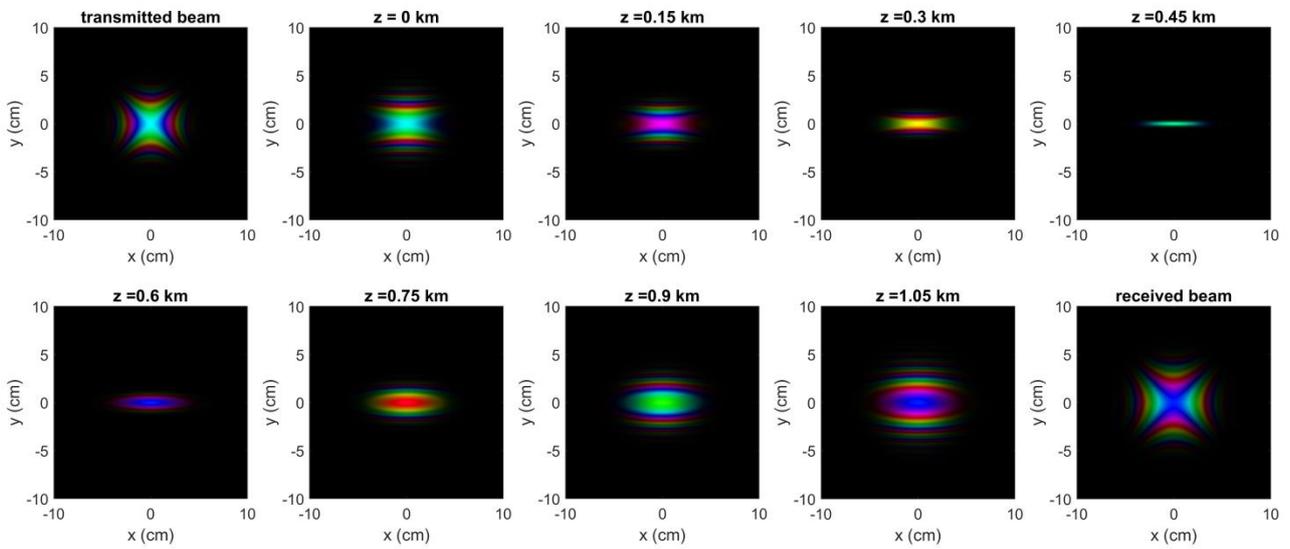

**Figure S7.** Free-space transmission over 1.05 km of a multipole-phase beam of order $m=+2$, with phase strength $\alpha_0 = 5.625 \cdot 10^{-9} \mu m^{-2}$, orientation angle $\vartheta_0 = 0$, beam waist $w_0 = 3.44 cm$ (transmitted beam). A quadratic lens term with focal length $f = 1.05 km$ is applied to the beam before propagation ($z=0$). At a distance $f$, after applying a quadratic-phase correction, a multipole-phase beam of the same order is retained (received beam), with $\alpha_1 = 3.974 \cdot 10^{-9} \mu m^{-2}$, rotation angle $\vartheta_1 = \pi/2$, beam waist $w_1 = 4.09 cm$.

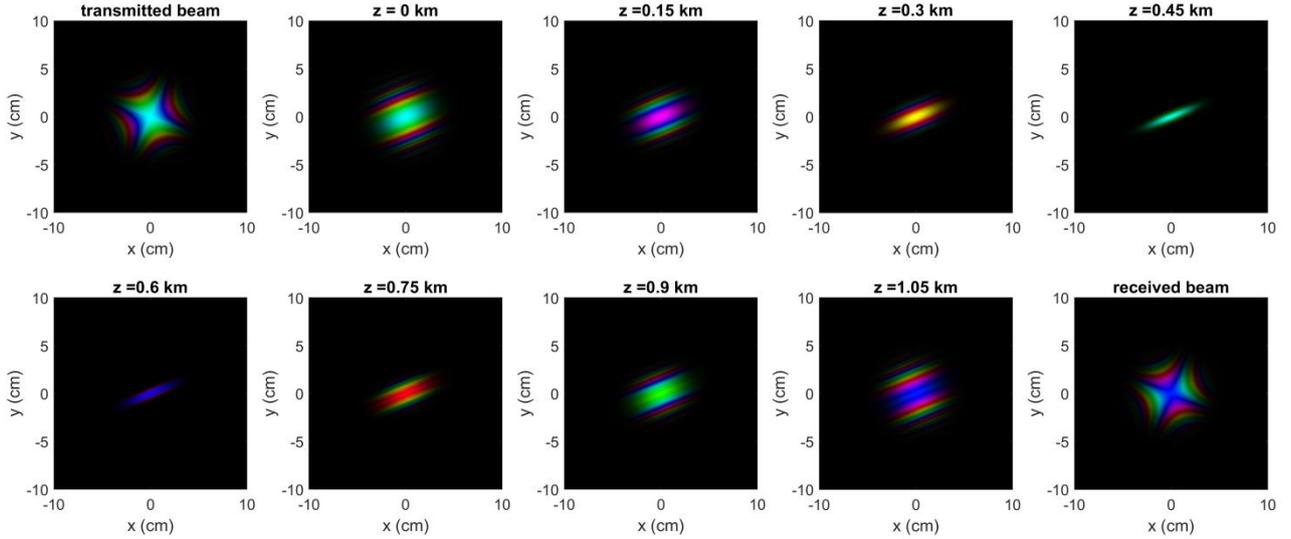

**Figure S8.** Free-space transmission over 1.05 km of a multipole-phase beam of order $m = +2$, with phase strength $\alpha_0 = 4.500 \cdot 10^{-9} \mu m^{-2}$, orientation angle $\vartheta_0 = \pi/8$, beam waist $w_0 = 3.44 cm$ (transmitted beam). A quadratic lens term with focal length $f = 1.05 km$ is applied to the beam before propagation ($z = 0$). At a distance $f$, after applying a quadratic-phase correction, a multipole-phase beam of the same order is retained (received beam), with $\alpha_1 = 4.798 \cdot 10^{-9} \mu m^{-2}$, rotation angle $\vartheta_1 = 5\pi/8$, beam waist $w_1 = 3.33 cm$.